# A Fiber Bundle Model of Systemic Risk in Financial Networks

Soumyajyoti Biswas[1]  and Bikas K. Chakrabarti[2]

## Abstract

Failure statistics of banks in the US show that their sizes are highly unequal (ranging from a few tens of thousands to over a billion dollars) and also, they come in `waves' of intermittent activities. This motivates a self-organized critical picture for the interconnected banking network. For such dynamics, recent developments in studying the inequality of the events, measured through the well-known Gini index and the more recently introduced Kolkata index, have been proved to be fruitful in anticipating large catastrophic events. In this chapter we review such developments for catastrophic failures using a simple model called the fiber bundle model. We then analyze the failure data of banks in terms of the inequality indices and study a simple variant of the fiber bundle model to analyze the same. It appears, both from the data and the model, that coincidence of these two indices signal a systemic risk in the network.

## 1. Introduction

Systemic Risk in the context of financial markets refers to the risks imposed by the network interlinks (e.g., in an interdependent bank network) in the market, where the failure of a single component or a number of them (a bank or a cluster of banks) can cause a cascading failure, which could potentially bankrupt or bring down the entire system or market. Although such failures are extensively addressed and studied in the financial literature (see e,g. [1] for a collection of reviews), straight forward modellings of these cascading failures in such networks are still absent.

In physics, however, there are precise and very well studied models, called the fiber bundle models (FBM). Indeed, the model follows an old well known and extensively studied model of materials failures, like fracture, earthquake etc., (see e.g.,

[1] Soumyajyoti Biswas
Department of Physics, SRM University - AP,
Department of Computer Science and Engineering, SRM University - AP,
Andhra Pradesh 522240, India, e-mail: soumyajyoti.b@srmap.edu.in
[2] Bikas K. Chakrabarti
Saha Institute of Nuclear Physics, Kolkata 70064, India;
Economic Research Unit, Indian Statistical Institute, Kolkata 700108, India
e-mail: bikask.chakrabarti@saha.ac.in

[2-7]), where a bundle of fibers or strings of different strength collectively supports the load on (or hanging from) the bundle. Failure of any individual fiber increases the load share on each of the remaining fibers and that may induce failure of the next weakest fiber and so on until the remaining fibers are strong enough to support the load on the bundle. Or else, all the fibers break and a catastrophe (like fracture of the material) occurs.

Another similar model for traffic jams may also be compared. Here a local jam in a part of the traffic network increases the (diverted) traffic load on the available free roads, some of which in turn gets jammed due to this increased load, leading to further increase in traffic load on the surviving links or roads. These (local failures) may not still lead to a total failure of the traffic network if the surviving roads can sustain the traffic load of the system (city traffic) and the system survives. Otherwise, it will lead eventually to a total jamming in the city lasting for the day and comes back to normalcy in the night when traffic load decreases (see e.g., [8] for a simple analytically tractable model of such traffic jams).We intend to propose here a similar FBM for systemic risk of financial networks  where the net financial `stress' W is assumed to get equally shared among $N(t)$ banking units, where $N(t = 0) = N$ (initially). Each such unit is assumed to have some threshold of its own and that threshold has uniform distribution. When the distributed load on any one goes beyond its capacity, it fails and the load per surviving units increase from $W/N(t)$ to $W/[N(t) - 1]$  and further failure may occur due to this increased load. Representing the fraction of surviving financial units at any time $t$ by $U(t)$, one can write (see e.g., [3, 5]) the dynamical equation $-\frac{dU}{dt} = [U^2 - U + \sigma]/U$; with $\sigma = I/N$, giving the surviving fraction $U(t) = U^* + (\sigma_c - \sigma)^{1/2}$, with $\sigma_c = 1/4$ near but before the collapse of the system, when all the fixed point fraction $U^*$ of the units fail, simultaneously. One can study [9], both analytically and numerically, the inequality indices for the avalanche sizes as the FBM approaches the critical point (tuning the stress $\sigma$ towards $\sigma_c$). It may be mentioned at this point that such an FBM is tuned externally (by changing the stress level $\sigma$). One can also consider self-tuned FBM systems, like the sand-piles where the external drive need not be tuned and dynamically stays put in the Self-Organized Critical (SOC) state (see e.g., [10]). In such SOC versions of FBM, the inequality indices for the avalanche distributions show some universal behavior [11].

We will attempt to extract and correlate the above-mentioned universal critical (SOC in particular) behavior with those from real data. Particularly, the data from the Federal Deposit Insurance Corporations (FDIC) [12] for bank failures in the US between 1934-2023 reveal that such failures often happened in clusters of events and a large event is preceded by smaller events. This is similar to what is seen in the failure models discussed above. Therefore, we will attempt to find statistical similarities and potential precursory signals, specifically from the point of view of the inequalities of the failure events that has recently been proved to be useful in indicating imminent catastrophic events.

In what follows, we first review avalanche dynamics in the fiber bundle model, discuss the dynamics of the model modified for the purpose of banking network. Then we define the inequality indices for the avalanche dynamics and review the analytical expressions for the same. We then measure the inequality indices as obtained from the real data (FDIC data) and compare it with those obtained from the model.

## 2. Avalanches in the Fiber Bundle Model

As mentioned before, the fiber bundle consists of $N$ elements or fibers which collectively support (through ``rigid platforms'' at both top and hanging end) a load $W = N\sigma$ and failure threshold ($\sigma_{th}$) of the fibers are assumed to be different for different fibers in the bundle. Initially, when a stress or load per fiber ($\sigma$) is applied, the fibers having failure threshold ($\sigma_{th}$) lower than the applied stress breaks immediately and the entire load then gets redistributed among the surviving fibers. In case of the Equal Load Sharing or ELS FBM considered here, the load is uniformly redistributed. The dynamics stops either when there is no fiber having threshold within this increased load per fiber or when all the $N$ fibers have failed. For simplicity, we assume here the threshold distribution of the fibers to be uniform within the range 0 to 1 (normalized). If $U_t(\sigma)$ denotes the fraction of surviving fibers at time (load redistribution iteration) $t$, then the further broken fiber fraction $U_{t+1}$ is given by load per fiber at that time $\sigma_{tn} = W/NU_t$. Hence,

$$U_{t+1} = 1 - \sigma_t = 1 - \frac{\sigma}{U_t} \tag{1}$$

At fixed point ($U_{t+1} = U_t = U^*$),

$$U^*(\sigma) - \frac{1}{2} = (\sigma_c - \sigma)^{1/2}; \sigma_c = 1/4. \tag{2}$$

If the order parameter is defined as $O \equiv U^*(\sigma) - U^*(\sigma_c)$ then,

$$O = (\sigma_c - \sigma)^\beta; \beta = 1/2. \tag{3}$$

One can also consider the failure susceptibility $\chi$, defined as the change of $U^*(\sigma)$ due to an infinitesimal increment of the applied stress $\sigma$

$$\chi = \left| \frac{dU^*(\sigma)}{d\sigma} \right| = \frac{1}{2}(\sigma_c - \sigma)^{-\gamma}; \gamma = \frac{1}{2}. \tag{4}$$

Employing Josephson's identity in the Rushbrooke equality, we get $2\beta + \gamma = d\nu = 3/2$, with $\nu$ being the correlation length exponent for the ELS FBM).

### 2.1 Inequality of avalanches: Lorenz function

Up to a given time i.e., step of load increment in the simulation of the FBM, the series of the avalanches can be arranged in the ascending order of their sizes. Then the Lorenz function [13] $L(p,t)$ can be calculated by the cumulative fraction of the avalanche mass (sum of all avalanche sizes) coming from the $p$ fraction of the smallest avalanches up to time $t$. Note that if all avalanches were of equal sizes, then the Lorenz function would be a diagonal line from the origin (0,0) to (1,1). This line is called the equality line (see Fig. 1). Since the avalanches are, in general, not of equal sizes, the Lorenz function in non-linear, always staying below the equality line and monotonically increasing, with the constraints that $L(0,t) = 0$ and $L(1,t) = 1$ for any $t$. The area in between the equality line and the Lorenz function, therefore, is a measure of the inequality in the avalanche sizes (the shaded area in Fig. 1). The ratio of this area and that under the equality line ($\frac{1}{2}$ by construction) is called the Gini index $g$ [14].

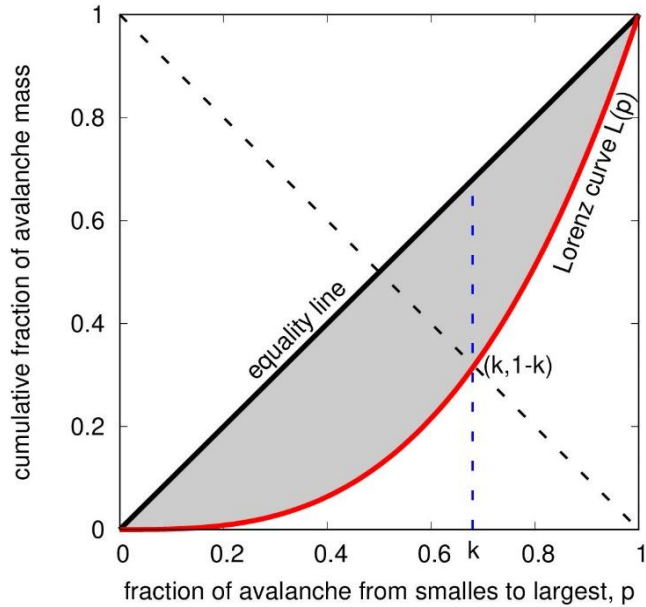

*Figure 1: The schematic diagram shows the Lorenz function (in red) and the equality line (black diagonal). The shaded area, as mentioned in the text, is therefore a measure of*

*inequality of the events concerned. That area (normalized by maximum inequality) is called the Gini index. On the other hand, the intersection of the opposite diagonal and the Lorenz function gives the Kolkata index (k), denoting 1-k fraction of the largest events accounting for k fraction of the total damage.*

On the other hand, the ordinate value of the crossing point of the opposite diagonal (straight line between (0,1) to (1,0)), gives the value of the Kolkata index $k$ [15], which gives the fraction $1 - k$ of the total number of avalanches that collectively account for the $k$ fraction of the total avalanche mass up to that time. It is a generalization of the Pareto's law [16] that says about 80% of `attempts' account for 20% of `successes'. It was previously noted that in the case of breakdown in the FBM, at the terminal point $t = t_f$, $k$ approaches a value close to $0.62 \pm 0.03$, irrespective of disorder strengths and system sizes.

It was also noted elsewhere that for a broad class of systems, predominately of socio-economic nature, the early time variations of $g(t)$ and $k(t)$ follow a linear relation $k(t) = 1/2 + 0.37g(t)$. This relation is empirical and seen in data. It turns out that such linearity is also observed in the simulations for the FBM.

### 2.1.1 Calculating the Gini and Kolkata indices for Fiber Bundle Model

Given its mean-field nature, it is possible to calculate the avalanche size distribution and the critical threshold (load at which the system collapses) for the fiber bundle model. Therefore, by extension, it is also possible to calculate the Gini and the Kolkata indices as follows:

We start from the definition of susceptibility, as mentioned above: $\chi = |dU^*(\sigma)/d\sigma| \propto (\sigma_c - \sigma)^{-1/2}$. Physically, this implies that a small change in the load, results in a `large' response in terms of breaking of fibers, particularly when $\sigma \to \sigma_c$. Naturally, the `responses' i.e., the avalanches are highly unequal and can thus be quantified using the indices mentioned above.

In doing so, one can write the Lorenz function, for all avalanches until the catastrophic breakdown at $\sigma = \sigma_c$, as

$$L_f(p) = \frac{\int_0^{p\sigma_c}(\sigma_c - \sigma)^{-1/2}d\sigma}{\int_0^{\sigma_c}(\sigma_c - \sigma)^{-1/2}d\sigma} = 1 - \sqrt{1 - p}. \tag{5}$$

From the above Lorenz function, the Gini index at the point of catastrophic failure can be calculated as

$$g_f = 1 - 2 \int_0^1 L_f(p)dp = 1/3 \tag{6}$$

The Kolkata index then is

$$1 - k_f = 1 - \sqrt{1 - k_f} \tag{7}$$

which gives

$$k_f = \frac{\sqrt{5}-1}{2} \approx 0.618 \tag{8}$$

This is what is numerically seen as well. Note that the values of these indices are not dependent on the critical point. This means that these are universal quantities, which will be seen as long as the particular power-law divergence is seen. Therefore, monitoring the inequality of responses can act as a good indicator of imminent failure. Indeed in sand-piles or self-organized fiber bundle models (see e.g., [11]) the inequalities (as measured by the indices $g$ and $k$) of the toppling or avalanche sizes universally show $g = k \simeq 0.86$ just preceding the arrival or following the departure of the SOC dynamical states.

It is this universal character of the inequality indices that we wish to utilize in failure statistics of interconnected network such as banking and the systemic risks, as observed from historic data and fiber bundle like model.

## 3. Avalanches in Bank Collapse

The banking system can be viewed as an interconnected network, where an overall financial stress can cause failures of individual banks and, if continued, can lead up to a systemic risk of catastrophic failure (such as the housing bubble of 2008-09).

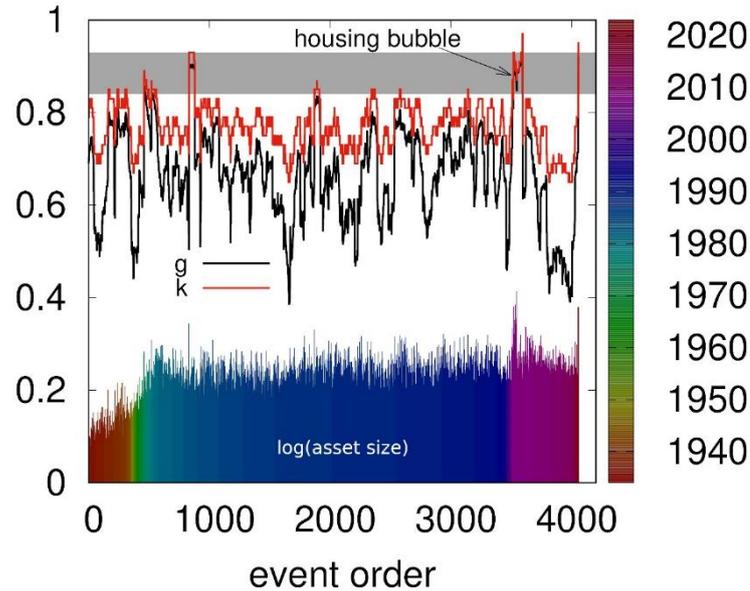

*Figure 2: The inequality of the bank failure sizes (shown in log of sizes S) are measured over a moving window of 50 events. The colors indicate the year of a particular failure. It is seen that the inequality indices g and k come very close to each other during the early 80s, then around 2009 and then very recently! The crossing of g and k is highlighted at the time of the housing bubble. The grey shaded region indicates the values for which the crossing might happen and therefore can be indicative of imminent large failure.*

From the data available with the Federal Deposit Insurance Corporation (FDIC) in the US, the bank failure sizes can be seen to be highly unequal. Indeed, in several cases, there have been events of growing sizes, leading up to major or `catastrophic' failures. In what follows, we will analyze the data for bank failures in terms of the inequality of their sizes and model the dynamics using simple fiber bundle like structure.

### 3.1 Inequality indices for bank collapse data

The FDIC website [12] lists bank failure (in assets size of deposits $S$) data from 1934 to March, 2023. The sizes of the failures, measured in terms of the deposits, vary widely (from several tens of thousands to over a billion dollars). It is, therefore, natural to measure the inequality in those failure data.

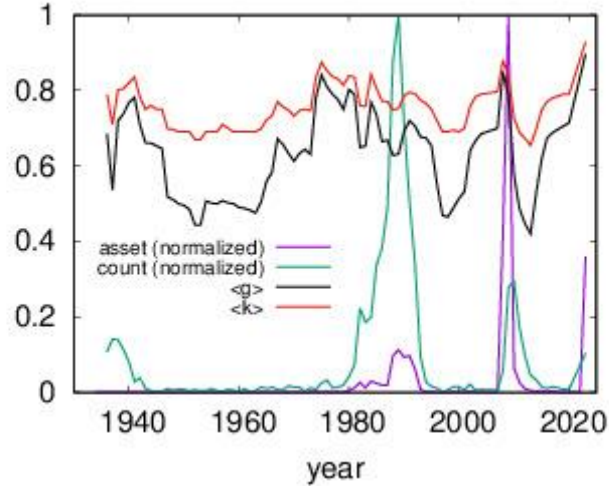

*Figure 3: The yearly average of g and k values noted in Fig. 2, denoted by ⟨g⟩ and ⟨k⟩, are shown along with the yearly cumulative failure sizes (normalized by the maximum size) and the number of failures (also normalized by the maximum number). This shows, until very recently, there were three major `waves' of bank failures in the US, in 30s, 80s and the housing bubble of 2008-09. In all such cases, ⟨g⟩ and ⟨k⟩ came very close to each other.*

In Fig. 2, the inequality of the bank failures were measured over a moving window of 50 events. This ensures that only the events of the recent past influence the values of $g$ and $k$. It is seen that $g$ and $k$ cross a couple of times and most prominently near the housing bubble of 2008-09. Also, it is noted that the current failures are again leading $g$ and $k$ values close to each other.

In Fig. 3, the yearly average of the inequality indices was shown with the (normalized) cumulative sizes of failure each year and the (normalized) cumulative number of failures each year. It is seen that there have been three periods (until very recently), where major banking failures happened in the US -- in 1930s, 1980s and the housing bubble of 2008-09. In all those cases, ⟨g⟩ and ⟨k⟩ came very close to each other.

### 3.1.1 A model for bank failure

Informed by the intermittent nature of the bank failure data, we mention a minimal model inspired by the fiber bundle model mentioned before.

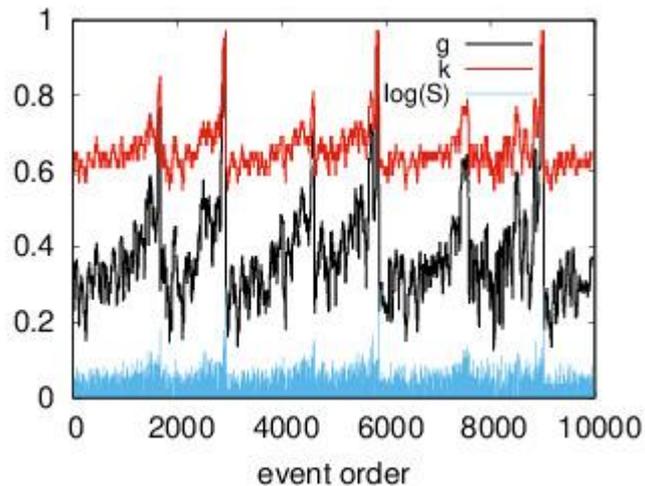

*Figure 4: The inequality indices g and k are studied for the SOC fiber bundle model discussed in section 3.1.1. The failure avalanches (S) in assets are also indicated in the log scale. The crossing of g and k within a range of value 0.82 ± 0.05 occurs near the major cascading failures in the bank network.*

The model consists of $N$ nodes (fibers), each having a failure threshold, which denotes the load carrying capacity. If the load exceeds the threshold, the node is broken, and the load is shared equally by the surviving nodes.

It is realistic to assume that the failure thresholds have a lower cut-off. For our purposes, we take the lower cut-off to be 0.5 and the thresholds are uniformly distributed between (0.5,1.5). The system is then loaded until the minimum threshold i.e., 0.5. The dynamics of the model then continues as follows: At a stable configuration, one node is randomly selected, and its threshold is set to zero. This means that the node collapses and the load is redistributed equally among all the surviving nodes. This may lead to further failure and an avalanche can start. The avalanche will continue until all nodes have loads lower than their respective thresholds. When such a stable configuration is reached, one more node is randomly selected, and its threshold is set to zero. Note that the system now is already `stresses' following the previous avalanche. So, in general an avalanche of higher size is expected. This process continues, until a macroscopic fraction of the nodes is eliminated (we set the threshold at 0.2 or 20% of nodes here). Following this, the broken nodes are restored with randomly chosen thresholds in the same range as before. In this way, the dynamics can continue as long as needed, with avalanches with different sizes. A similar form of the model was studied for the self-organized state of power grids.

## 4. Summary and conclusions

Systemic failures in banking networks have been modeled here employing the Fiber Bundle Models or FBMs [2] for which the failure dynamics have been studied extensively in the recent physics literature in the context of fracture (in inhomogeneous materials) and earthquakes (see e.g., [3-6]) and traffic jams (see e.g., [8],[17,18]). This mapping here helps us, compared to some earlier studies (see e.g., [19]), to provide some precise analysis of systemic or collective failure dynamics of the networks in the so called mean field (or long-range load reallocation) limit, as discussed in section 2. We also extend here mapping to the self-organizing critical (SOC) limit, and study the universal SOC behavior of bank failures in section 3.

As discussed earlier, in a road network if the traffic load in one road goes beyond its capacity that road gets jammed, and the diverted traffic load gets redistributed to its link roads. This extra load share may induce jams in one or more of these link roads which, in turn, may induce further jams and may result in cascading failure of the entire network. An exactly solvable FBM model, in the mean field limit [8], has been given in sec. 2. As the load $\sigma$ gets increased, the steady state fraction $U^*(\sigma)$ of the intact (on service) banks (fibers/roads) decreases following a critical or power law behavior (see eqn. (3)) for the uniform load level or stress on the links $\sigma$ in the range $0 \leq \sigma < \sigma_c$, where $\sigma_c = 1/4$ (normalized), and a discontinuous or catastrophic failure collapse (from the steady state bank or link fraction $U^*(\sigma_c) = 1/2$ to $U^*(\sigma_c) = 0$) occurs at the critical load level $\sigma_c = 1/4$. Next, we discuss in section 3, how the cascading bank collapse in the US (data for the period 1930-2020 from [12]) can be comprehended using such fiber bundle models. In particular, in the SOC limit [10,11] of such FBM models, where the broken fibers are replaced by intact fibers having breaking thresholds from the same distribution (see sub-section 3.1.1), one can search further precursors (see Figs. 2 and 3). Here the systemic bank collapses are detected by searching the event points (or times) where the Gini ($g$) and Kolkata ($k$) indices (defined in Fig. 1) in the failure avalanche statistics come very close ($g = k$ in the range $0.82 \pm 0.05$). It may be noted here again that $k = 0.8$ corresponds to Pareto's 80-20 law [15, 16] of inequality. It is also interesting to note that inequality (measured in terms of Gini index) in accumulated gain or wealth through a Minority Game, reaches the highest unequal point, with $g$ slightly above 0.8, at the point of maximum cooperation (minimum dispersion) [20].

## Acknowledgements

BKC is grateful to the Indian National Science Academy for their Senior Scientist Research Grant. Numerical analyses were performed using HPCC Surya at SRM

University - AP. The authors are grateful to Parongama Sen for discussions and critical reading of the manuscript.

# References

[1] Abergel F, Chakrabarti BK, Chakraborti A and Ghosh A (Eds.) (2013) Econophysics of Systemic Risk and Network Dynamics, New Economic Windows Series, Springer.

[2] Pierce FT (1926) The Weakest Link. J. Textile Inst. 17:355 [https://doi.org/10.1080/19447027.1926.10599953]

[3] Pradhan S, Hansen A, and Chakrabarti BK (2010) Failure Processes in Elastic Fiber Bundles. Rev. Mod. Phys. 82:499.

[4] Kawamura H, Hatano T, Kato N, Biswas S, and Chakrabarti BK (2012) Statistical Physics of Fracture, Friction, and Earthquakes. Rev. Mod. Phys. 84:839.

[5] Biswas S, Ray P and Chakrabarti BK (2015) Statistical Physics of Fracture, Breakdown, and Earthquake: Effects of Disorder and Heterogeneity, Wiley-VCH.

[6] Hansen A, Hemmer PC, Pradhan S (2015) The fiber bundle model: modeling failure in materials, Singapore: Wiley-VCH.

[7] Chakrabarti BK (2017) Story of the development in statistical physics of fracture, breakdown and earthquake: A personal account. Rep. Adv. Phys. Sci. 1:1750013 [doi:10.1142/S242494241750013X]

[8] Chakrabarti BK (2006) A fiber bundle model of traffic jams. Physica A: Statistical Mechanics and its Applications 372:162-166 [https://doi.org/10.1016/j.physa.2006.05.003].

[9] Diksha, Kundu S, Chakrabarti BK and Biswas S (2023) Inequality of avalanche sizes in models of fracture. Phys. Rev. E 108:014103.

[10] P. Bak, How Nature Works: The Science of Self-Organized Criticality, Springer-Verlag, New York (1996)

[11] Manna SS, Biswas S and Chakrabarti BK (2022) Near universal values of social inequality indices in self-organized critical models, Physica A: Statistical Mechanics and its Applications 596:127121.

[12] Federal Deposit Insurance Corporations (FDIC), https://www.fdic.gov/bank/statistical/index.html

[13] Lorenz MO (1905) Methods of measuring the concentration of wealth. Publications of the American Statistical Association 9:209–219.

[14] Gini CW, Variabilitá e Mutabilitá: Contributo allo Studio delle Distribuzioni e delle Relazioni Statistiche; Cristiano Cuppini: Bologna, Italy (1912)

[15] Ghosh A, Chattopadhyay N, Chakrabarti BK (2014) Inequality in societies, academic institutions and science journals: Gini and k-indices. Physica A: Statistical Mechanics and its Applications 410:30–34.

[16] Pareto V (1897) Cours d'Economie Politique, Lausanne, Rouge.

[17] Zheng JF, Zhao XM and Fu BB (2008) Extended fiber bundle model for traffic jams on scale free networks, International Journal of Physics C 19:1727-1735.

[18] Batool A, Danku Z, Pal G and Kun F (2022) Temporal evolution of failure avalanches of the fiber bundle model on complex networks. Chaos: An Interdisciplinary J. Nonlin. Sc.32:063121.

[19] Lorenza J, Battiston S and Schweitzer F (2009) Systemic risk in a unifying framework for cascading processes on networks, European Physical Journal B, 71: 441–460. DOI: 10.1140/epjb/e2009-00347-4.

[20] Ho KH, Chow FK, Chau HF (2004) Wealth inequality in the minority game, Phys. Rev. E, 70:066110.